\documentclass[aip,pop,reprint,groupedaddress, amsmath, amssymb, showpacs]{revtex4-1}

 \usepackage{gensymb}
 \usepackage{subcaption}
 \usepackage{caption}
\newcommand*{\rom}[1]{\expandafter\@slowromancap\romannumeral #1@}
\newcommand{\vc}{\mathbf}
\usepackage{graphicx}
\usepackage{color}
\begin{document}

\title{A Generalized Boltzmann Kinetic Theory for Strongly Magnetized Plasmas with Application to Friction}


\author{Louis Jose and Scott D. Baalrud}
\affiliation{Department of Physics and Astronomy, University of Iowa, Iowa City, Iowa 52242, USA}


\date{\today}

\begin{abstract}

Coulomb collisions in plasmas are typically modeled using the Boltzmann collision operator, or its variants, which apply to weakly magnetized plasmas in which the typical gyroradius of particles significantly exceeds the Debye length. Conversely, O'Neil has developed a kinetic theory to treat plasmas that are so strongly magnetized that the typical gyroradius of particles is much smaller than the distance of closest approach in a binary collision. Here, we develop a generalized collision operator that applies across the full range of magnetization strength. To demonstrate novel physics associated with strong magnetization, it is used to compute the friction force on a massive test charge. In addition to the traditional stopping power component, this is found to exhibit a transverse component that is perpendicular to both the velocity and Lorentz force vectors in the strongly magnetized regime, as was predicted recently using linear response theory. Good agreement is found between the collision theory and linear response theory in the regime in which both apply, but the new collision theory also applies to stronger magnetization strength regimes than the linear response theory is expected to apply in.

\end{abstract}




\maketitle





\section{Introduction}


It is well known that magnetic fields influence the transport of energy, momentum and particles in plasmas. However, most plasmas are weakly magnetized in the sense that the gyroradius of particles is much larger than the Debye length. In this situation, the influence of the magnetic field during Coulomb collisions is negligible. In contrast, if the magnetic field is sufficiently strong that the gyroradius of particles is smaller than the Debye length, the trajectories of colliding particles are influenced by the magnetic field during a Coulomb collision. In this strongly magnetized regime, kinetic theory must account for gyromotion at the microscopic collision scale; i.e., within the collision operator.  This is a formidable challenge in part because there is no analytic solution of the two-body problem in the presence of a magnetic field. Despite this difficulty, understanding the influence of strong magnetization on transport processes remains interesting both from the point of view of fundamental physics as well as its importance in many experiments, such as magnetic confinement fusion~\cite{Aymar_2002}, non-neutral plasmas~\cite{PhysRevLett.68.317}, ultracold neutral plasmas~\cite{PhysRevLett.100.235002}, magnetized dusty plasmas~\cite{Thomas_2012}, trapped antimatter~\cite{fajans2020plasma} and naturally occurring plasmas in planetary magnetospheres~\cite{2004jpsm.book..593K}. This work presents a generalized plasma kinetic theory that applies to conditions spanning from weak to strong magnetization regimes.

Regimes that characterize the influence of magnetization on transport can be identified by comparing the gyroradius~($r_c = \sqrt{k_B T/m}/\omega_c $) with the Debye length~($\lambda_D = \sqrt{k_B T/ 4 \pi e^2 n} $), Landau length times $\sqrt{2}$~($r_{L} =\sqrt{2}e^2/k_B T$) and Coulomb collision mean free path~($\lambda_{\textrm{col}}$)~\cite{PhysRevE.96.043202}. Here, we apply the generalized kinetic theory to compute the friction force on a massive test charge moving in a magnetized one-component plasma~(OCP). The magnetized OCP is described by two dimensionless parameters: the Coulomb coupling strength, $\Gamma = (e^2/a)/(k_BT)$, where $a=3/(4\pi n)$ is the Wigner-Seitz radius and the magnetic field strength $\beta = \lambda_D/r_c = \omega_c/\omega_p$, where $\omega_c=eB/mc$ is the gyrofrequency and $\omega_p = \sqrt{4\pi e^2n/m}$ is the plasma frequency.  For a weakly coupled plasma~($\Gamma <1$), the four possible transport regimes~\cite{PhysRevE.96.043202} are summarized in table~\ref{tab:table1}.

\begin{table}[b]
\caption{\label{tab:table1}%
Four different transport regimes of a weakly coupled one-component plasma
}
\begin{ruledtabular}
\begin{tabular}{l l l}
1   & Unmagnetized plasma 		  	& $r_c > \lambda_{\textrm{col}}$\\
2   & Weakly magnetized plasma 		& $\lambda_D < r_c < \lambda_{\textrm{col}}$\\
3   & Strongly magnetized plasma    & $r_L<r_c<\lambda_D$\\
4   & Extremely magnetized plasma   & $r_c<r_L<\lambda_D$ \\
\end{tabular}
\end{ruledtabular}
\end{table}

In traditional plasma kinetic theory, the transport properties are obtained from the Boltzmann equation: $\partial_t f + \vc{v}~ \cdot~ \partial_X f +\frac{e}{m}[\vc{E}+\vc{v}/c \times \vc{B}]\cdot \partial_v f = \mathcal{C}$, in which the interactions between particles occurring at microscopic scales are described by the collision operator ($\mathcal{C}$). The traditional theory~\cite{ferziger1972mathematical} assumes that the gyromotion of the interacting particles occurs at a length scale that is much larger than the Debye scale volume within which binary collisions occur. As a result, the collision operator is independent of the magnetic field, and the theory applies to the unmagnetized and weakly magnetized regimes. O'Neil has developed a kinetic theory~\cite{o1983collision} that treats the opposite limit, in which particles are so strongly magnetized that the gyroradius is smaller than the Landau length. This theory considers single component plasmas and makes use of a property of adiabatic invariance in a collisions that applies in the extremely magnetized regime (region 4). Other techniques to treat strongly magnetized plasmas include Fokker-Planck equations~\cite{cohen2019collisional,cohen2018fokker,montgomery1974fokker,montgomery1974magnetic,dubin2014parallel}, binary collisions using perturbation theory~\cite{dong2013effects,dong2013temperature}, generalized Lenard-Balescu theories~\cite{rostoker1960kinetic} and guiding-center approximations~\cite{PhysRevLett.62.51,pitaevskii2012physical,PhysRevLett.79.2678}.

In this work, we present a generalized Boltzmann kinetic theory that treats all four magnetization regimes. In this approach, a microscopic collision volume is defined in the coordinate space where the collisions occur. Colliding particles enter the collision volume, interact and scatter out of it, having exchanged momentum and energy inside the collision volume. The traditional Boltzmann collision operator is obtained in the weak magnetic field limit and O'Neil's collision operator is obtained as the high magnetic field limit.

The generalized collision operator predicts that strong magnetization gives rise to new transport properties that are not predicted by the traditional Boltzmann equation. To illustrate this, it is used to compute the friction force on a massive test charge moving through a background plasma in the presence of an external magnetic field of varying strength. The common understanding based on the traditional kinetic theory is that the friction force acts antiparallel to the velocity vector of the test charge~\cite{nersisyan2014interactions}. The friction force is also commonly referred to as stopping power. The stopping power determines the energy loss of the projectile and has implications for the energy deposition of fusion products~\cite{Sigmar_1971}.

The generalized kinetic theory predicts that a qualitatively new effect arises when the plasma is strongly magnetized. Instead of being aligned antiparallel to the velocity vector, the friction force is predicted to shift, gaining a transverse component that is perpendicular to the velocity vector in the plane formed by the velocity and magnetic field vectors. This transverse component alters the trajectory and the stopping distance of the projectile~\cite{lafleur2019transverse,lafleur2020friction}. It's sign is predicted to change if the test charge speed is either faster or slower than approximately the thermal speed of the plasma species with which it predominately interacts. It acts to increase the gyroradius of fast particles and to decrease the gyroradius of slow particles. Since the transverse force is perpendicular to the velocity, it does not decrease the energy of the projectile and is unnoticed in the conventional way of obtaining the friction force from the energy loss of the projectile~\cite{nersisyan2014interactions, PhysRevE.79.066405, PhysRevE.67.026411, PhysRevE.61.7022,cereceda2005stopping}. The existence of this transverse friction force was recently predicted using a different approach to kinetic theory based upon a linear dielectric response formalism~\cite{lafleur2019transverse} and was confirmed using first-principles MD simulations~\cite{PRL_David}. The assumptions inherent to the linear response approach limit its applicability to regimes 1-3 in table 1. Good agreement between the two approaches is demonstrated throughout this range of conditions, but the new collision based approach extends the range of applicability because it also applies in the extremely magnetized regime (region 4). 

The outline of the paper is as follows. In Sec. II, the generalized collision operator is derived and the traditional Boltzmann collision operator and O'Neil's collision operator are obtained in the appropriate asymptotic limits. Section III describes the theoretical formulation and numerical evaluation of the friction force on a single test charge. Section IV discusses the results of these computations in regimes of weak, strong and extreme magnetization, along with a comparison of theories. Section V gives a qualitative description of the transverse force from the binary collision perspective.
 
\section{ Generalized Collision operator}
Derivations of the Boltzmann equation begin from a general description of the dynamics of $N$ interacting particles, but then apply a series of approximations to focus on average quantities of interest, and to make the problem tractable by invoking properties of the dilute gas (or plasma) limit. Here, we follow a traditional derivation due to Grad~\cite{harris2004introduction,grad1958principles,cercignani2013mathematical}, insofar as it applies to arbitrary magnetization. This includes defining reduced distribution functions, and making use of the dilute limit to justify binary collisions, the molecular chaos approximation~\cite{harris2004introduction}, and local collisions that happen at microscopic space and time scales. The departure from the traditional derivation comes about by not making approximations of a certain geometry for the collision volume that are justified only in the absence of strong magnetization, and by accounting for the Lorentz force when computing the binary collision dynamics inside that volume. This leads to a more general, but more computationally intensive, kinetic equation. It reduces to either the traditional result or O'Neil's result in the appropriate limits.
 
Grad's derivation begins from Liouville's theorem~\cite{ferziger1972mathematical}, which describes the phase space evolution of the $N$-particle distribution function ($f^{(N)}(\vc{r}_1,\vc{r}_2 \cdots \vc{r}_N,\vc{v}_1, \vc{v}_2 \cdots \vc{v}_N,t)$), but this description is a complex computational problem because of the large number of degrees of freedom brought by the huge number of particles in the system. As a way to reduce the computational complexity of the problem and to focus on physical processes of interest, reduced distributions are defined by integrating a subset of the degrees of freedom. This results in the BBGKY hierarchy. These equations are not closed because the evolution equation for the $n$ particle distribution ($f^{(n)}(\vc{r}_1,\vc{r}_2 \cdots \vc{r}_n,\vc{v}_1, \vc{v}_2 \cdots \vc{v}_n,t)$) contains the $n+1$  particle distribution.
 
	The Boltzmann equation can be derived from the BBGKY hierarchy by applying the closure $f^{(3)} = 0$, which drops triplet correlations. Although this method is accurate for particles interacting via short-range potentials, such as neutral gases, it leads to an infrared divergence in plasmas due to the long-range nature of the Coulomb interaction~\cite{ferziger1972mathematical}. This is usually corrected by introducing an ad hoc cutoff to the impact parameter at the Debye length to model Debye screening. Recent work~\cite{baalrud2019mean} using a new closure of the BBGKY hierarchy has shown that expanding about the deviations of correlations from their equilibrium values, 
\begin{equation}
\Delta f^{(3)} = f^{(3)} - f_o^{(3)} f^{(2)}/f_o^{(2)}
\end{equation}
rather than in terms of the strength of correlations, $f^{(3)}$, ensures that the exact equilibrium properties are maintained at all orders of the hierarchy, including screening. Here, $f_o$ is the equilibrium distribution function. This expansion shows that binary collisions occur via the potential of mean force~\cite{PhysRevLett.110.235001,baalrud2014extending}, rather than the bare Coulomb potential. The potential of mean force asymptotes to the Debye-H\"{u}ckel potential in the weakly coupled limit
\begin{equation}
\phi(r) = \frac{e_1 e_2}{r} e^{-r/\lambda_D},
\end{equation}
which is the only case we consider here. Here, $r = |\vc{r}| = |\vc{r}_1 - \vc{r}_2|$ is the distance between the particles and $e_1$, $e_2$ are the charges of the particles. We make use of this new closure, but otherwise follow Grad's~\cite{harris2004introduction,grad1958principles,cercignani2013mathematical} derivation of the Boltzmann equation. 

\begin{figure}[h]  
\centerline{\includegraphics[width = 2.0in]{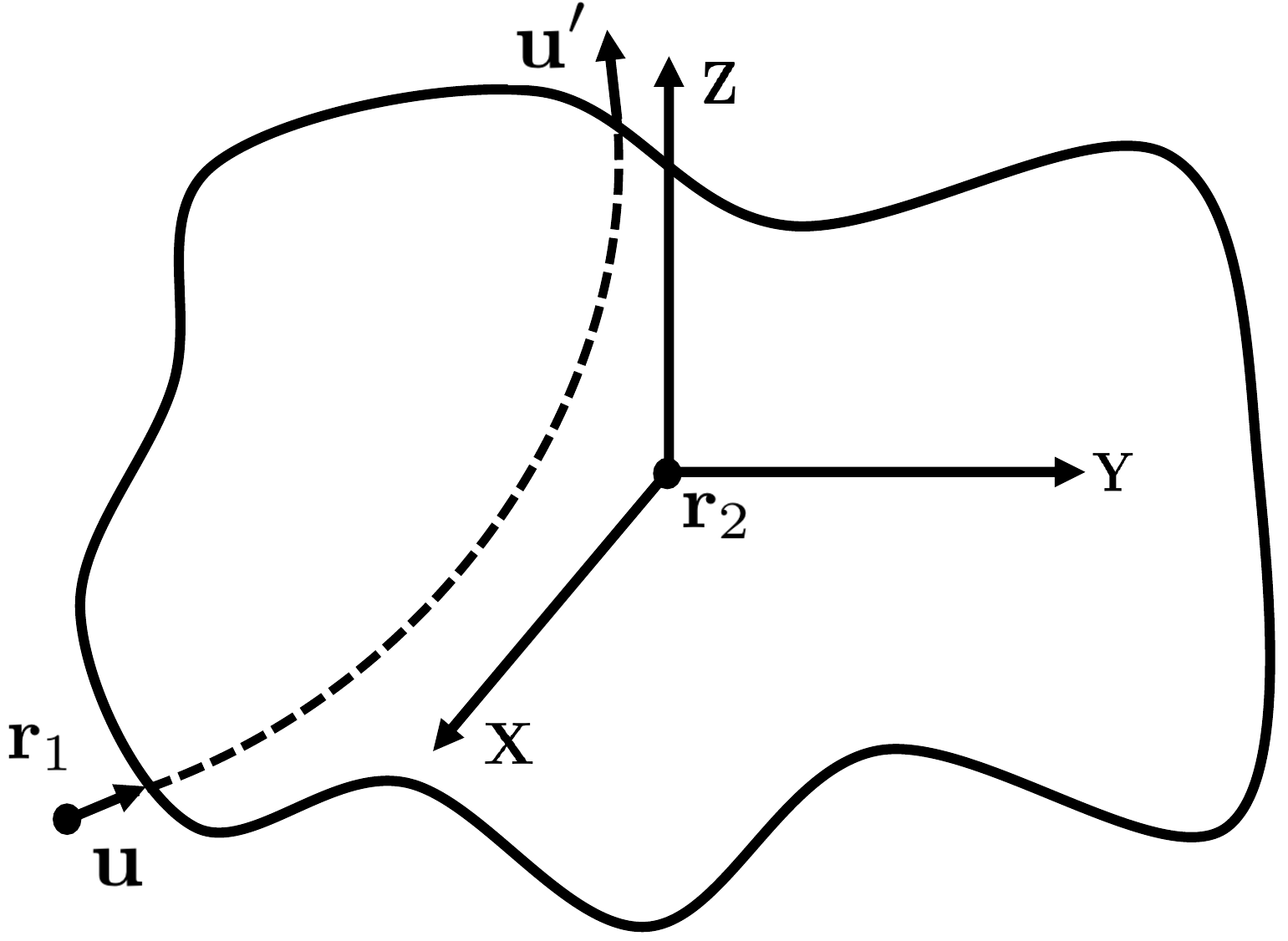}}
\caption{Illustration of a collision volume surrounding particle 2 ($\vc{r}_2$) during an interaction with particle 1 ($\vc{r}_1$).}
  \label{fig:vol}
\end{figure}

In this approach, the coordinate space is divided into a microscopic volume where the collisions occur and an outside region where no collisions occur. In this way, the particle distribution function is divided into two: a short range component in which the microscopic collisions occur and a long range component (truncated distribution function) which is uniform on the microscopic length scale. In the small collision volume limit, the truncated distribution function is the observable particle distribution function. The collision operator obtained after following these arguments is~\cite{baalrud2019mean}
  
\begin{equation}
\mathcal{C} =  \int d^3 \vc{v}_2 \oint_S ds \,  \vc{u} \cdot \vc{\hat{s}} \, f^{(2)}(\vc{r}_1,\vc{r}_2,\vc{v}_1,\vc{v}_2,t),
\end{equation}
where $\vc{u} = \vc{v}_1 - \vc{v}_2$ is relative velocity between the colliding particles and $f^{(2)}(\vc{r}_1,\vc{r}_2,\vc{v}_1,\vc{v}_2,t)$ is the two particle distribution function. The surface integral is defined by the small region of space in which the two particles interact (collision volume). Here, $ds$ is the infinitesimal area on the surface and $\vc{\hat{s}}$ is the unit normal of the area element. The collision volume can be visualized in the relative frame with the coordinate system fixed to particle $2$ with particle $1$ entering the collision volume as shown in Fig. \ref{fig:vol}.

Depending on the sign of $\vc{u} \cdot \vc{\hat{s}}$, the surface integral is split into two terms, representing contributions from two surfaces  $S_{+} (\vc{u} \cdot \vc{\hat{s}} > 0)$ and $S_{-} (\vc{u} \cdot \vc{\hat{s}} < 0)$.
Points on $S_{+}$ correspond to particles moving away from each other (postcollision state) and points on $S_{-}$ correspond to particles moving towards each other (precollision state). Assuming binary collisions occurring via the mean force within the interaction volume, the second order BBGKY equation with the $\Delta f^{(3)} = 0$ truncation provides a solution whereby $ f^{(2)}(\vc{r}_1,\vc{r}_2,\vc{v}_1,\vc{v}_2,t)$ is a constant along the two particle trajectory. Thus, for points on the $S_+$ surface, $f^{(2)}$ can be replaced with the post collision coordinates: $f^{(2)}(\vc{r}_1,\vc{r}_2,\vc{v}_1,\vc{v}_2,t)\rightarrow f^{(2)}(\vc{r}_1^\prime,\vc{r}_2^\prime,\vc{v}_1^\prime,\vc{v}_2^\prime,t)$, where the variables with prime($'$) represent postcollision states~\cite{harris2004introduction,grad1958principles}. On making these changes, the collision operator becomes 
\begin{eqnarray}
\mathcal{C} =   \int d^3 \vc{v}_2 \int_{S+} ds \, |\vc{u} \cdot  \vc{\hat{s}}|  f^{(2)}(\vc{r}_1^\prime,\vc{r}_2^\prime,\vc{v}_1^\prime,\vc{v}_2^\prime,t) \notag \\ 
- \int d^3 \vc{v}_2 \int_{S-}ds \,| \vc{u} \cdot  \vc{\hat{s}} | f^{(2)}(\vc{r}_1,\vc{r}_2,\vc{v}_1,\vc{v}_2,t).
\end{eqnarray}

In order to get an explicit form of the collision operator, the integrals can be written over the same integration range by the change of variable: $\vc{\hat{s}}  \rightarrow -\vc{\hat{s}} $ in the first integral, changing  $S_+$ to $S_-$. Further, we make the assumption of molecular chaos ($Stosszahlansatz$): $f^{(2)}(\vc{r}_1,\vc{r}_2,\vc{v}_1,\vc{v}_2,t) =  f^{(1)}(\vc{r}_1,\vc{v}_1,t)f^{(1)}(\vc{r}_2,\vc{v}_2,t)$, where $f^{(1)}(\vc{r},\vc{v},t)$ is the one particle distribution function~\footnote{Considering collisions in strongly magnetized one-component plasmas, Dubin has argued that strong magnetization causes particles to recollide multiple times~\cite{PhysRevLett.79.2678}. Such multi-body interactions violate the molecular chaos approximation. Further research will be required to understand the role of recollisions in the friction force problem considered in this work, or to adapt the generalized collision operator to incorporate recollision dynamics.}. We also assume that the collision volume is small. This is justified by the short timescale of the collision compared to the larger timescale of evolution of $f^{(1)}(\vc{r},\vc{v},t)$. In this limit of a local collision $\vc{r}_1^\prime$,$\vc{r}_2^\prime$ and $\vc{r}_2$ can be approximated by $\vc{r}_1$~\cite{harris2004introduction,grad1958principles}. On making these approximations, a generalized collision operator is obtained 
 \begin{eqnarray} \label{gco}
\mathcal{C} =   \int d^3 \vc{v}_2 \int_{S_-} ds \, |\vc{u} \cdot  \vc{\hat{s}}|  (f_1'f_2' - f_1f_2),
 \end{eqnarray}
in which the following abbreviated notations have been applied 
 \begin{eqnarray}
& f_1' \equiv  f^{(1)}(\vc{r}_1,\vc{v}_1^\prime,t),\, \, \,\, \, \, \, \,  f_2' \equiv  f^{(1)}(\vc{r}_1,\vc{v}_2^\prime,t) , \notag \\
& f_1 \equiv  f^{(1)}(\vc{r}_1,\vc{v}_1,t), \, \, \, \, \, \, \, \, \, f_2 \equiv  f^{(1)}(\vc{r}_1,\vc{v}_2,t) . \notag
\end{eqnarray}

Equation~(\ref{gco}) is the expression for the generalized collision operator. This result has been obtained in many prior works during the path to derive the traditional Boltzmann equation. The novelty here is to evaluate Eq.~(\ref{gco}) directly, rather than proceed to simplify it by invoking arguments associated with either the weak or extreme magnetization limits. In order to evaluate this expression, the post collision velocities ($\vc{v}_1^\prime$ and $\vc{v}_2^\prime$) need to be evaluated. This involves solving the two body dynamics of the colliding particles inside the collision volume for the initial velocities ($\vc{v}_1$ and $\vc{v}_2$).

Equation (\ref{gco}) is a 5-D integral: 3-D velocity space volume and a 2-D surface in the coordinate space. The surface integral encloses a small region where collisions occur that is determined by the range of the potential of mean force; in this case, the Debye length. The integral can be viewed as summing over all possible configurations in which particle 1 enters the collision volume and interacts with particle 2, weighted by the postcollision and the precollision velocity distributions. The surface integral counts all the possible orientations in which the particle enters the collision volume and the velocity integral counts all the possible velocities of particle 2. Limiting the surface integral to the surface $S_{-}$ makes sure only the precollision states are counted. 
 
 Since we focus on weakly coupled plasmas for which the potential of mean force is the Debye-H\"{u}ckel potential, the collision volume is characterized by the Debye length scale. The equations of motion for two charged particles with masses $m_1$ and $m_2$ and charges $e_1$ and $e_2$ interacting in a uniform magnetic field $\vc{B}$ within the collision volume are

 \begin{eqnarray}\label{main}
m_{1} \frac{d\vc{v}_{1}}{dt} &=& -\nabla_{\vc{r}_1}\phi(r)+ e_1\Big(\frac{\vc{v}_{1}}{c}\times\vc{B}\Big) \\
m_{2} \frac{d\vc{v}_{2}}{dt} &=& -\nabla_{\vc{r}_2}\phi(r) +e_2\Big(\frac{\vc{v}_{2}}{c}\times\vc{B}\Big) .
\end{eqnarray}
 
 Since the Debye-H\"{u}ckel potential depends only on the distance between the particles, it is useful to change the variables to the center of mass,
\begin{eqnarray}\label{tra}
\mathbf{R}&=&\frac{m_{1} \mathbf{r}_{1}+m_{2} \mathbf{r}_{2}}{m_{1}+m_{2}} ,  \\
\mathbf{V}&=&\frac{m_{1} \mathbf{v}_{1}+m_{2} \mathbf{v}_{2}}{m_{1}+m_{2}} , 
\end{eqnarray} 
and the relative frame, $\vc{r}$ and $\vc{u}$. Under this transformation, the equations of motion for the center of mass and the relative velocities are
\begin{eqnarray}
&&(m_1+m_2)\frac{d\vc{V}}{dt} = m_{12} \Big(\frac{\vc{u}}{c} \times \vc{B}\Big) \Big( \frac{e_1}{m_1}-\frac{e_2}{m_2} \Big) \notag \\
 && \, \, \, \, \, \, \, \, \, \,\, \, \, \, \, \, \, \, \, \,\, \, \, \, \, \, \, \, \, \,\, \, \, \, \, \, \, \, \, \, \, \, \, \, \, \, + (e_1+e_2) \Big(\frac{\vc{V}}{c} \times \vc{B}\Big), \label{eoms1} \\
&&m_{12}\frac{d\vc{u}}{dt} = -\nabla \phi(r)+ m_{12}^2 \Big(\frac{\vc{u}}{c} \times \vc{B}\Big) \Big(\frac{e_1}{m_1^2}+\frac{e_2}{m_2^2}\Big)  \notag \\ 
&& \, \,\, \,\, \,\, \, \, \, \, \, \, \, \, \, \, \, \, \, \, \, \, \, \, \, \,  +m_{12}  \Big(\frac{\vc{V}}{c} \times \vc{B}\Big) \Big(\frac{e_1}{m_1}-\frac{e_2}{m_2}\Big) .\label{eoms2}
\end{eqnarray}
These equations of motion describe how the precollision states transform to the postcollision states within the collision volume. The initial positions of the colliding particles in the relative frame correspond to the surface of the collision volume. For initial conditions corresponding to a precollision state, $\vc{u} \cdot \vc{\hat{s}} < 0$, the equations of motion are solved within the collision volume to obtain the postcollision velocities associated with the location at which the particles leave the collision volume. However, as indicated in Fig. \ref{fig:vol}, the precise shape of the collision volume can be arbitrary. The only relevant characteristic is that it must be significantly larger than the range of the potential of mean force~\cite{baalrud2019mean}. As long as this condition is met, both the precollision and postcollision states correspond to a condition in which there is negligible interaction between the particles. In practice, choosing a collision volume that is much larger than the range of the potential of mean force increases the computational cost associated with evolving the particle trajectories, but it does not alter the momentum exchanged, which is the input to the collision operator resulting from trajectory calculation. 

Although the general case does not require a specified volume, certain limits create symmetries that can be used to simplify the problem through the specification of a definite collision volume. We next consider two limiting cases: the traditional Boltzmann collision operator and O'Neil's collision operator. 



	
\subsection{Weakly magnetized limit: Boltzmann equation  }	
In the unmagnetized and weakly magnetized regimes, the effect of the magnetic field in the collision volume is negligible. The dynamics of the colliding particles are modeled using the Debye-H\"{u}ckel potential alone. The spherical symmetry of the Debye-H\"{u}ckel potential suggests that a sphere is an appropriate collision volume.
\begin{figure}[h]  
\centerline{\includegraphics[width = 1.85in]{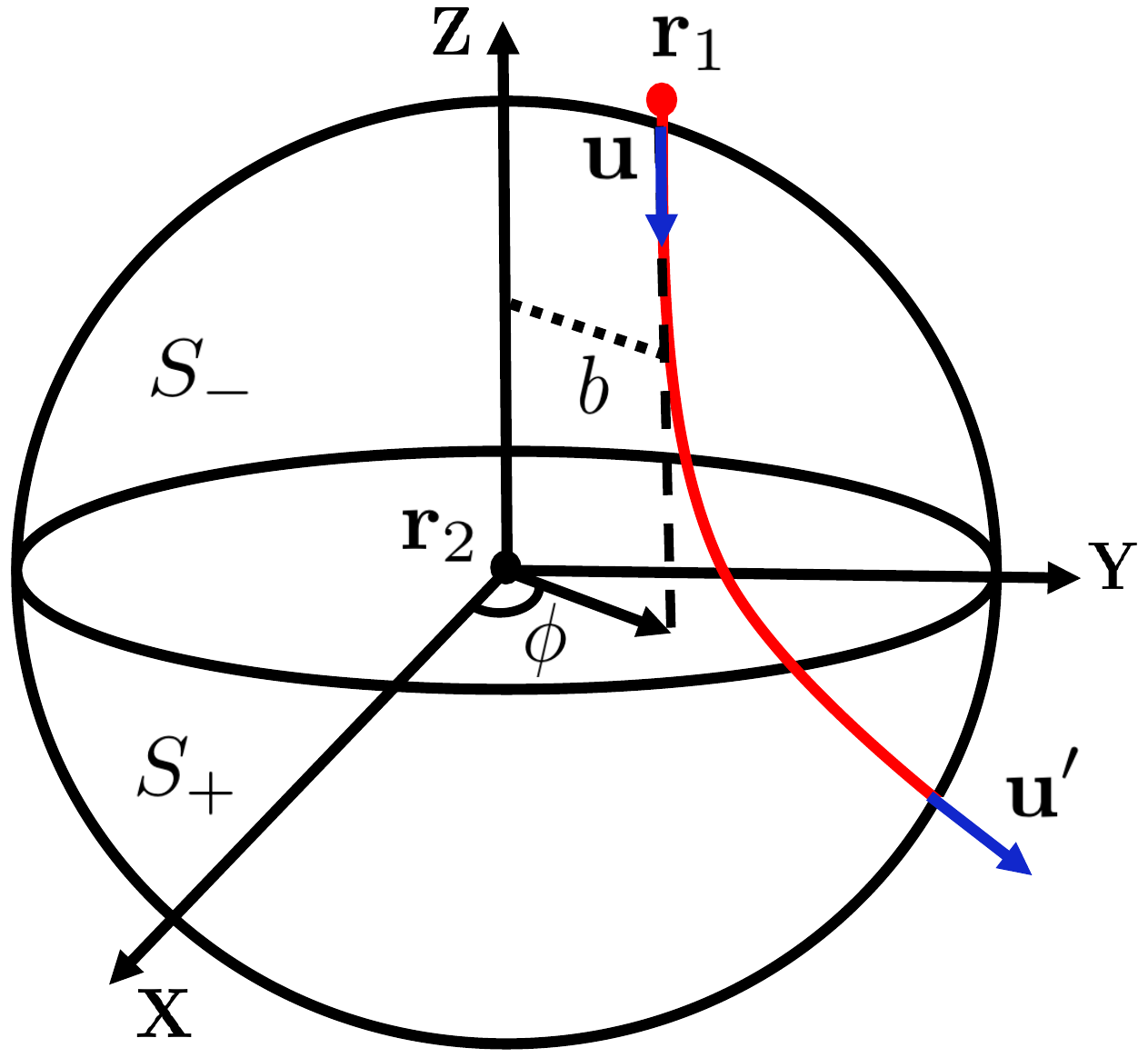}}
\caption{A spherical interaction volume around particle 2 ($\vc{r}_2$) in the presence of a repulsive interaction with particle 1 ($\vc{r}_1$). The disk surface is perpendicular to the precollision relative velocity ($\vc{u}$). Each point on the hemisphere ($S_-$) can be projected to a point on the disk as shown making a one to one correspondence. }
  \label{fig:sph}
\end{figure}

The usual form of the collision operator can be obtained by introducing a plane perpendicular to the relative velocity $\vc{u}$ of the colliding particles and intersecting the sphere along a diameter with the coordinate system centered at particle $2 $ ($\vc{r}_1$) as shown in Fig. \ref{fig:sph}. The $S_-$ surface is a hemisphere and the projection of it on this plane is a disk surface. Points on this disk surface have a one-to-one correspondence with the points on the hemisphere. This enables a transformation of the integration surface from the hemisphere to the disk surface. Using $b \, db \, d\phi$ as the area element of the polar coordinates on this disk, the collision operator can be recast as~\cite{harris2004introduction,ferziger1972mathematical}
\begin{equation}
\mathcal{C} =   \int d^3 \vc{v}_2 \int b \, db \, d\phi \, u  (f_1'f_2' - f_1f_2),
\end{equation}
where $u=|\vc{u}|$ and $b$ can be identified as the impact parameter from Fig. \ref{fig:sph}. Substituting $b \, db \, d\phi = \sigma d\Omega$, where $d\Omega = \sin\theta d\theta d \phi$ is the differential scattering cross section and $d\Omega$ is the solid angle, we get 
\begin{equation}\label{coulomb_ope}
\mathcal{C} =   \int d^3 \vc{v}_2 \int \sigma d\Omega \, u  (f_1'f_2' - f_1f_2).
\end{equation}
 This is the traditional Boltzmann collision operator. It is much simpler to evaluate than the general form of Eq.~(\ref{gco}) because in the absence of a Lorentz force inside the collision volume, classical mechanics provides a closed form expression for the differential scattering cross section from the scattering angle~\citep{ferziger1972mathematical} 
\begin{equation}
\sigma = \frac{b}{\sin \theta} \biggl| \frac{db}{d\theta} \biggr|
\end{equation}
where $\theta = \pi - 2\Theta$ and
 \begin{equation}
 \Theta = b \int_{r_0}^{\infty} dr \frac{1}{r^2}\bigg[ 1-\frac{b^2}{r^2} - \frac{2 \phi(r)}{m_{12}u^2} \bigg]^{-1/2} \label{Theta}
 \end{equation}
 is the scattering angle and $r_0$ is the distance of closest approach, obtained by finding the root of the denominator of the integrand. In this case, the problem reduces to solving the scattering angle integral (Eq.~(\ref{Theta})). This is much simpler than the general case of solving the equations of motion (Eqs. (\ref{eoms1}) and (\ref{eoms2})) of the colliding particles inside the collision volume, which are coupled ordinary differential equations.

\subsection{Extremely magnetized limit: O'Neil equation}
 O'Neil developed a Boltzmann-like collision operator~\cite{o1983collision} that accounts for the collisions between particles of a one-component plasma in the extremely magnetized regime (region 4). This was later used to calculate the temperature anisotropy relaxation rate of a non-neutral plasma, and the predicted relaxation rate was validated experimentally~\cite{o1985collisional,glinsky1992collisional}. In this subsection, we show that O'Neil's results can be obtained from the generalized collision operator in the extremely magnetized limit by choosing the collision volume to be a cylinder.

\begin{figure}[h]  
\centerline{\includegraphics[width = 1.5in]{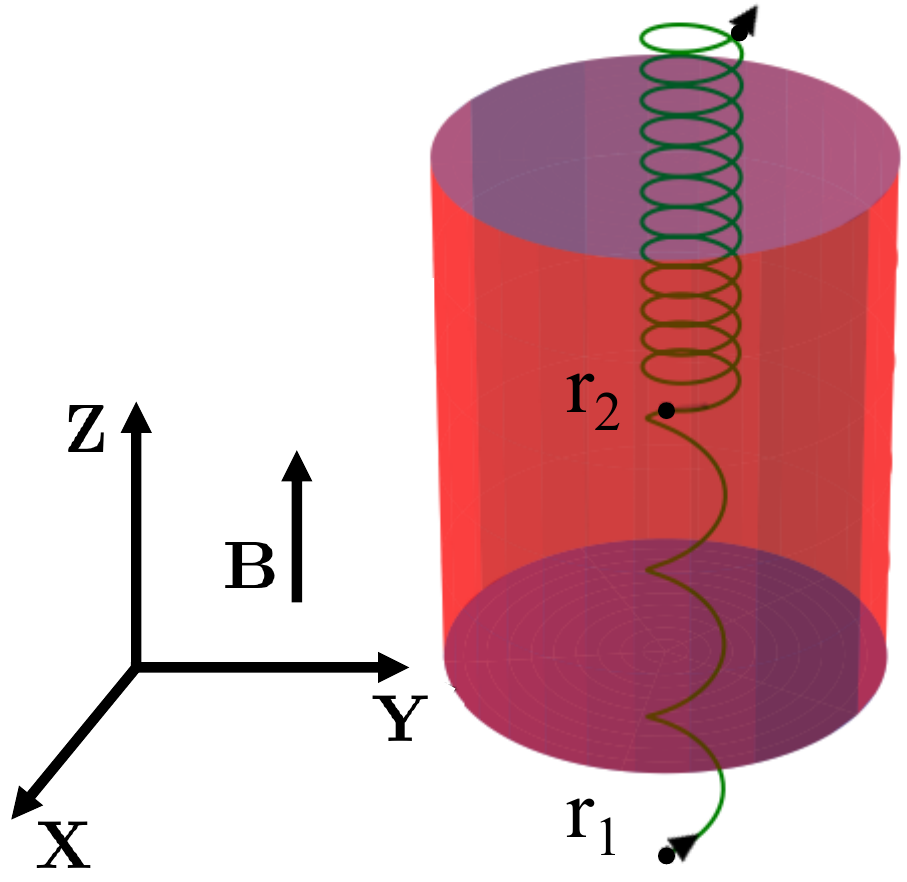}}
\caption {Cylindrical interaction volume around particle 2 ($\vc{r}_2$) during an interaction with particle 1 ($\vc{r}_1$) in the presence of a strong magnetic field.}
  \label{fig:cyl}
\end{figure}

The one-component plasma is a special case because when the charge-to-mass ratio is the same, the center of mass motion and relative motion are decoupled (Eqs. (\ref{eoms1}) and (\ref{eoms2})). The resulting equations of motion in the relative frame are equivalent to that of a charged particle in a uniform magnetic field scattered by the potential at the origin. The resulting trajectory is that of a helix before and after the collision, but where both the parallel and gyromotion can change due to the collision. The natural geometry characterizing this motion is a cylinder, as depicted in Fig. \ref{fig:cyl}.

For this geometry of the collision volume, Eq. (\ref{gco}) takes the form
 \begin{eqnarray}
 \mathcal{C} =   \int d^3 \vc{v}_2 \int \rho d\rho d\theta  \, |\vc{u} \cdot  \vc{\hat{z}}|  (f'_1f'_2- f_1f_2) \notag \\ 
  +  \int d^3 \vc{v}_2 \int \rho d\theta dz  \, |\vc{u} \cdot  \vc{\hat{\rho}}|  (f'_1f'_2- f_1f_2).  
 \end{eqnarray} 
The first term corresponds to collisions in which particle 1 enters through the circular surface (blue), and the second term corresponds to collisions in which particle 1 enters through the cylindrical surface (red) depicted in Fig. \ref{fig:cyl}. When the plasma is extremely magnetized, the gyroradius of particles is so small compared to the size of the collision volume ($\lambda_D$) that motion is restricted to remain close to the guiding centers and scattering perpendicular to the initial guiding centers is minimal. In this case, the contribution to the collision operator from the second term is negligible. In this limit 
\begin{equation}
\mathcal{C} =   \int d^3 \vc{v}_2 \int \rho d\rho d\theta  \, |\vc{u} \cdot  \vc{\hat{z}}|  (f'_1f'_2- f_1f_2).
\end{equation} 
This is O'Neil's collision operator for extremely magnetized plasmas~\cite{o1983collision}. As there is no closed-form solution of the equations of motion of colliding particles in the presence of a magnetic field, they are solved numerically to find the postcollision states of the particles, except in the asymptotic case of a very large magnetic field~\cite{glinsky1992collisional}. Even though the evaluation of the transport coefficients using O'Neil's theory is more difficult than the traditional Boltzmann theory, it is simpler than the general theory because the equations of motion are decoupled making the numerical calculation of the trajectories less computationally expensive.

\section{Friction force}

 The generalized collision operator can be used to compute the macroscopic transport properties of the plasma in all magnetization strength  regimes. To illustrate this, we compute the friction force acting on a massive projectile, taken to be a single test charge, moving through a magnetized one-component plasma. Understanding how friction is modified in the presence of a strong magnetic field is fundamentally important and it also has direct implications in many magnetized plasma experiments such as non-neutral plasmas~\cite{PhysRevLett.68.317}, ultracold neutral plasmas~\cite{PhysRevLett.100.235002}, magnetic confinement fusion~\cite{Aymar_2002} and naturally occurring plasmas in planetary magnetospheres~\cite{2004jpsm.book..593K}. It is also the fundamental process controlling macroscopic transport of momentum.

A projectile moving through the plasma is acted upon by friction in addition to the Lorentz force from the external magnetic field. The friction force is due to Coulomb collisions with the background plasma. In the unmagnetized and weakly magnetized regimes, the Boltzmann equation predicts that the friction force is antiparallel to the velocity of the projectile and is commonly known as the stopping power~\cite{baalrud2016effective}. It was recently predicted that a qualitatively new effect occurs in strongly magnetized plasmas: the friction force obtains a transverse component that is perpendicular to the velocity vector of the projectile, in the plane formed by $\vc{v}$ and $\vc{B}$~\cite{lafleur2019transverse}. This prediction was made using linear response theory, and was later tested using molecular dynamics simulations~\cite{PRL_David}. Here, we test the generalized collision operator by comparing with the predictions of linear response theory in the strongly magnetized regime. We also compute the friction force in the extremely magnetized regime, showing that the transverse component of the friction force exists, and is large, in this regime. This extends the regime of magnetization over which this phenomenon has been studied because the linear response theory from~\cite{lafleur2019transverse} is not expected to apply in the extremely magnetized regime due to a close-collision cutoff in that theory. 

\subsection{Theory}
Consider a massive projectile slowing down on a magnetized one-component plasma. Since the projectile is very massive compared to the background plasma, the gyromotion of the projectile happens at a larger spatial scale than the size of the collision volume. While the Lorentz force significantly influences the background plasma, it has a negligible influence on the massive test charge during the collision and can therefore be accurately excluded from the equations of motion for the test charge. The equations of motion from Eqs. (\ref{eoms1}) and (\ref{eoms2}) then reduce to 
\begin{eqnarray}
&&(m_1+m_2)\frac{d\vc{V}}{dt} = e \Big(\frac{\vc{V}}{c} \times \vc{B}\Big)-\frac{e \, m_{12}}{m_2} \Big(\frac{\vc{u}}{c} \times \vc{B}\Big)\label{main1} \\
&&m_{12}\frac{d\vc{u}}{dt} = -\nabla \phi(r)+ \frac{e \, m_{12}^2}{m_2^2} \Big(\frac{\vc{u}}{c} \times \vc{B}\Big)  \notag \\ 
&& \, \,\, \,\, \,\, \, \, \, \, \, \, \, \, \, \, \, \, \, \, \, \, \, \, \, \,  -\frac{e \, m_{12}}{m_2}  \Big(\frac{\vc{V}}{c} \times \vc{B}\Big) .\label{main2}
\end{eqnarray}
Here the charge on the massive projectile is taken to be the same as the charge of the background plasma particles ($e_1 = e_2 = e$). The friction force is $\vc{F} = \vc{R}^{12}/n_1$, where $n_1$ is the density of the projectile and $\vc{R}^{12}$ is the friction force density obtained by taking the momentum moment of the collision operator,
\begin{eqnarray}\label{fdensity}
\vc{R}^{12} = \int d^3 \vc{v}_1 \int d^3 \vc{v}_2 \int_{S_{-}} ds \, |\vc{u} \cdot  \vc{\hat{s}}| \notag \\
 m_1 \vc{v}_1 (f_1'f_2' - f_1f_2) .
\end{eqnarray}
Equation (\ref{fdensity}) can be simplified using the principle of detailed balance~\cite{glinsky1992collisional} ($\int d^3 \vc{v}_1 d^3 \vc{v}_2 ds \, |\vc{u} \cdot  \vc{\hat{s}}|  m_1 \vc{v}_1 f_1'f_2' = \int d^3 \vc{v}_1 d^3 \vc{v}_2 ds \, |\vc{u} \cdot  \vc{\hat{s}}| m_1 \vc{v}_1^\prime  f_1f_2$). The derivation of the principle of detailed balance typically relies upon the invariance of the collision dynamics under time-reversal and space inversion to find an inverse collision~\cite{reif2009fundamentals}. But for a system with an externally generated magnetic field, the time-reversal symmetry is no longer valid (the system has the time-reversal invariance only if the reversal of current direction that produces the magnetic field is accounted for) making it difficult to find an inverse collision. Nevertheless, the system of charged particles with an externally generated magnetic field is expected to follow detailed balance because many collisions can be lumped together to produce a result with the same consequence as an inverse collision. This argument is similar to that made in polyatomic gases, which are another system in which a inverse binary collision does not exist~\cite{cercignani1981h}. Although a proof has not been developed for the magnetized plasma case, as it has for the polyatomic gas~\cite{cercignani1981h}, we adopt the detailed balance relation as a postulate, as others have done~\cite{glinsky1992collisional}. 

After applying the detailed balance, the friction force density can be recast as
\begin{eqnarray}\label{r12}
\vc{R}^{12} = \int d^3 \vc{v}_1 \int d^3 \vc{v}_2 \int_{S_-} ds |\vc{u} \cdot  \vc{\hat{s}}| \notag \\
 m_1 (\vc{v}_1^\prime-\vc{v}_1)f_1f_2,   
\end{eqnarray}
where $\vc{v}_1^\prime$ is the postcollision velocity of the projectile, obtained by solving the equations of motion. Since the projectile is a single particle, its distribution is a Dirac delta function. The background plasma distribution is taken as a uniform Maxwellian distribution. Thus,
\begin{eqnarray}
f_1 &=& n_1 \delta^3(\vc{v}_1-\vc{v}_0), \notag \\
f_2 &=& \frac{n_2}{\pi^{3/2}v_T^3}\exp\bigg(\frac{-v_2^2}{v_T^2}\bigg), \notag
\end{eqnarray}
where $n_2$ is the density of the background plasma and $v_T = \sqrt{2 k_B T/m_2}$ is the thermal velocity of the background plasma. On making these substitutions we get,
\begin{eqnarray}
\vc{R}^{12} =  \frac{n_1 n_2 m_1}{\pi^{3/2}v_T^3}\int d^3 \vc{v}_2 \int_{S_-}  ds \, |\vc{u} \cdot  \vc{\hat{s}}|   \nonumber \\
(\vc{v}_1^\prime-\vc{v}_0)\exp\bigg(\frac{-v_2^2}{v_T^2}\bigg).
\end{eqnarray}

Three components of the friction force are obtained from the friction force density using the following definitions
\begin{subequations}
\begin{eqnarray}
F_v &=& \frac{\vc{R}^{12} \cdot \vc{{\hat{v}}}_0}{n_1}   , \label{sp} \\
F_{\times} &=& \frac{\vc{R}^{12} \cdot (\vc{{\hat{v}}}_0 \times \vc{\hat{n}}) }{n_1}  ,\label{tf}  \\
F_{n} &=& \frac{\vc{R}^{12} \cdot \vc{{\hat{n}}}}{n_1}, \label{vb} 
\end{eqnarray}
\end{subequations}
where $\vc{\hat{n}}$ is the unit vector perpendicular to $\vc{v}_0$ and $\vc{B}$ defined as $\vc{\hat{n}} = \vc{\hat{v}}_0 \times \vc{\hat{b}}/\sin \theta$, $\hat{b} = \vc{B}/|\vc{B}|$ is the unit vector in the direction of the magnetic field and $\theta$ is the angle between $\vc{v}_0$ and $\vc{B}$. Here, $-F_v$ is the stopping power, $F_{\times}$ is the transverse force and $F_{n}$ is the friction force component along the direction of the Lorentz force.	 
\subsection{Numerical evaluation}
The integrals for computing different components of the friction force are five-dimensional: three in the velocity space and two in the coordinate space. They were solved numerically using Monte Carlo integration. The computational difficulty is that the coupled differential equations (Eq. (\ref{main1}) and Eq. (\ref{main2})) describing the two-body interaction in a magnetic field must be solved numerically to compute the change in velocity of the projectile ($ \vc{v}_1^\prime-\vc{v}_0$) for each Monte Carlo integration point. Because the parameter-space is five-dimensional, a very large number of integration points is required for convergence. In our computations, the number ranged from $10^6$ to $10^8$. In order to solve the integrals numerically, the equations were first made dimensionless by normalizing the time with the plasma frequency, distance with the Debye length and velocity with the Debye length times the plasma frequency. Using the scaled variables,
\begin{widetext}
\begin{eqnarray}
\vc{R}^{12} &=& \frac{M k_B T n_1}{8\sqrt{6} \pi^{5/2}\Gamma^{3/2} \lambda_D}\int  \tilde{v}_2^2 \sin \theta_{v_2}  d\theta_{v_2} d\phi_{v_2} d\tilde{v}_2 
\int_{\tilde{S}_{-}} \tilde{R}_s^2  \sin \theta_{R_s}  d\theta_{R_s} d\phi_{R_s}  |\vc{\tilde{u}} \cdot  \vc{\hat{s}}|  (\vc{\tilde{v}}_1^\prime-\vc{\tilde{v}}_0)\exp\bigg(\frac{-\tilde{v}_2^2}{2}\bigg) . \label{spn} 
\end{eqnarray}
\end{widetext}
Here, the collision volume is taken as a sphere and the integrals are written in spherical polar coordinates for both velocity and space (recall that the shape of the collision volume is unimportant in the general theory, so long as it is large compared to the range of the interparticle force). Since the potential falls off exponentially on the Debye length scale, the radius of the sphere ($R_s$) is taken as 2.5 Debye lengths for computations in which the Coulomb coupling strength is $\Gamma = 0.1$ and 3.5 Debye lengths for $\Gamma = 0.01$. Here, M is the ratio of the mass of the projectile to that of the background plasma particle ($M=\frac{m_1}{m_2}$) and the variables with tilde (\~{}) on top represent scaled variables. 

A variety of Monte Carlo integration techniques are available to reduce the large number of sample points required in the integration routine. One common technique is the transformation method~\cite{glinsky1992collisional,press2007numerical}, but this requires an approximate analytic expression for the change in momentum of the projectile during a collision. As there is no known analytic expression for this problem, a different technique is desirable. Instead, we use an adaptive Monte Carlo integration technique - VEGAS~\cite{press2007numerical,PETERLEPAGE1978192,peter_lepage_2020_3647546}. In this method, the integration variables are recast in an attempt to make the integrand a constant and a Monte Carlo integration is performed. These two steps are iterated several times. The algorithm uses the information about the integrand from one iteration to optimize the change of variable for the next iteration.

 Friction force curves were obtained by evaluating the integrals (Eqs. (\ref{sp}), (\ref{tf}) and (\ref{vb})) based upon trajectory calculations with initial conditions selected from the adaptive Monte Carlo algorithm. Each point in this 5-D integral corresponds to an initial velocity of the background particle and an initial position in the relative coordinates
\def\B{
\begin{bmatrix}
    \tilde{R}_s \sin \theta_{R_s} \cos \phi_{R_s} \\
    \tilde{R}_s \sin \theta_{R_s} \sin \phi_{R_s}\\
    \tilde{R}_s \cos \theta_{R_s}
\end{bmatrix}}
\def\C{
\begin{bmatrix}
    \tilde{v}_2 \sin \theta_{v_2} \cos \phi_{v_2} \\
    \tilde{v}_2 \sin \theta_{v_2} \sin \phi_{v_2} \\
    \tilde{v}_{2} \cos \theta_{v_2}
\end{bmatrix}}
\begin{equation}
\vc{\tilde{r}}_i =\B  
\end{equation}
\begin{equation}
\vc{\tilde{v}}_{2i} = \C
\end{equation}
and the initial velocity of the projectile $\vc{\tilde{v}}_{1i}$ is taken as $\vc{\tilde{v}}_0 $. The unit normal vector is $\vc{\hat{s}}=\vc{\tilde{r}}_i/\tilde{R}_s$. Using Eq. ~(\ref{tra}), $\vc{\tilde{v}}_0 $ and $\vc{\tilde{v}}_{2i}$ were transformed to the relative and the center of mass coordinates. For initial states that satisfy $\vc{\tilde{u}} \cdot  \vc{\hat{s}} < 0$, the two particle equations of motion (Eq. (\ref{main1}) and Eq. (\ref{main2})) were solved using the 'DOP 853' method in the Python scipy library~\cite{2020SciPy-NMeth}. The trajectory calculations were stopped when the particle crossed the collision volume, i.e, $|\vc{\tilde{r}}| > \tilde{R}_s$ and the change of projectile velocity ($\vc{\tilde{v}}_1^\prime-\vc{\tilde{v}}_0$) was calculated. Twenty iterations of the VEGAS grid adaptation and the integral estimate were made. The final result and error was obtained by taking the weighted average of the last 10 iterations with the weight chosen to be the inverse of the variance of each of those iterations. Results from evaluation of the friction force in different transport regimes are discussed in the following section.

\section{Results}

In this section, we discuss the results from calculation of the friction on the projectile and compare it with the traditional Boltzmann theory and the linear response theory. The traditional Boltzmann theory is valid for the unmagnetized and weakly magnetized regimes and predicts that the stopping power is the only non-zero component of  the friction force. The expression for the stopping power when the interaction is modeled using the Debye-H\"{u}ckel potential in the traditional Boltzmann collision theory is~\cite{baalrud2016effective}
\begin{eqnarray} \label{clmb_sp}
F_v =  &&\frac{-n_2 m_{12} v_T}{2 \sqrt{\pi} |\vc{v}_0|^2} \int  du u^2 \sigma^{(1)}(u) \Big[ e^{-(u-|\vc{v}_0|)^2/v_T^2 } \,\,\,\,\,\,\,\,\,\,\,\,\,\,\,\,\, \nonumber \\
 &&\Big(\frac{2 u|\vc{v}_0|}{v_T^2} -1 \Big)+e^{-(u+|\vc{v}_0|)^2/v_T^2 }  \Big(\frac{2 u|\vc{v}_0|}{v_T^2} +1 \Big) \Big],
\end{eqnarray}
where 
\begin{equation}\label{crosssect}
\sigma^{(1)} = 4 \pi \int_{0}^{\infty} b db \cos^2\Theta(b,u) 
\end{equation}
 is the momentum-transfer scattering cross section and $\Theta(b,u)$ is the scattering angle (Eq. (\ref{Theta})).

	Linear response theory is valid in all the transport regimes except the extremely magnetized regime. It computes the friction force from the induced electric field associated with the wake generated by the movement of projectile in the plasma. The result is~\cite{lafleur2019transverse}
\begin{equation}
\vc{F} = -\frac{e^2}{2\pi^2} \int d^3 k   \frac{\vc{k}}{k^2} \text{Im}\bigg\{ \frac{1}{\hat{\epsilon} (\vc{k},\vc{k}\cdot \vc{v})}\bigg\},
\end{equation}
where $\vc{k}$ is wave vector and  
\begin{eqnarray}
\hat{\epsilon} (\vc{k},\vc{k}\cdot \vc{v}) \, &&= 1 + \frac{1}{k^2 \lambda_D^2} \bigg[ 1+ \frac{\vc{k}\cdot \vc{v} }{|k_{||}|v_T} \exp  \bigg(\frac{-k_\perp^2 v_T^2}{2 \omega_c ^2} \bigg) \bigg] \nonumber \\
&&\Bigg[\sum_{n = - \infty}^{\infty}  I_n \bigg(\frac{k_\perp^2 v_T^2}{2 \omega_c ^2} \bigg) Z \bigg( \frac{\vc{k}\cdot \vc{v} - n\omega_c}{|k_{||}|v_T} \bigg) \Bigg]
\end{eqnarray}
is the linear dielectric response function of the plasma. Here,  $Z$ is the plasma dispersion function~\cite{fried1961plasma}, $I_n$ is the $n$th order modified Bessel function of the first kind, and $k_{||} $ and $k_\perp$ are parallel and perpendicular components of the wave vector with respect to the direction of the magnetic field. Since the linear response theory does not account for the short range collisions, it would lead to a logarithmic divergence~\cite{nicholson1983introduction}. This is typically avoided by choosing a high wave number cut off  for the $k$ integral, $k_{\textrm{max}} = m_{12} (v_T^2 + v_0^2)/ e^2$, which is approximately the inverse of distance of closest approach~\cite{ichimaru2018statistical}.

\subsection{Unmagnetized and weakly magnetized plasma}

\begin{figure} [!htb] 
\centerline{\includegraphics[width = 1.05\columnwidth]{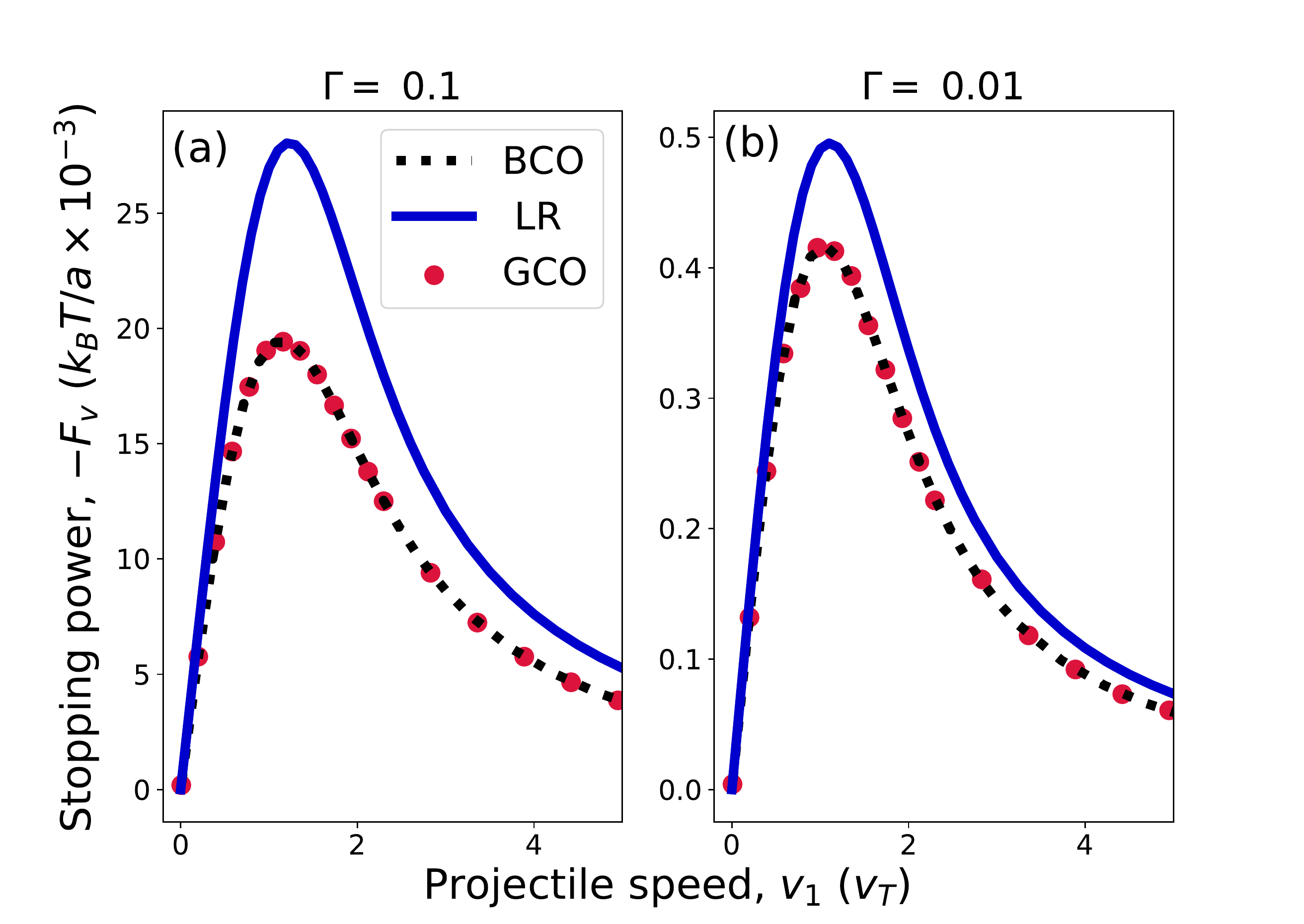}}
\caption {Stopping power ($-F_v$) of a massive projectile ($M=1000$) in a light background plasma with the coupling strengths (a) $\Gamma = 0.1$ and (b) $\Gamma =0.01$ using the generalized collision operator (GCO). Also shown are the predictions using the linear response theory (LR) and the traditional Boltzmann collision operator (BCO) }
  \label{fig:plt_gamma}
\end{figure}

In order to test the generalized collision operator, and our numerical implementation, we first compute the friction force in the unmagnetized and weakly magnetized plasma regimes and compare the results with the accepted results from the Boltzmann equation. Our computations are consistent with the expectation from the Boltzmann equation that, in this regime, only the stopping power component $F_v$ is non-zero~\cite{baalrud2016effective}. Figure \ref{fig:plt_gamma} compares the stopping power curve obtained using the generalized collision operator to the result of the traditional Boltzmann collision operator with the Debye-H\"{u}ckel potential as well as the results of linear response theory~\cite{ichimaru2018statistical} for $\Gamma=0.1$ and $\Gamma=0.01$. Since the influence of the magnetic field during collisions is negligible, the magnetic field was taken as zero in the equations of motion~(Eq. (\ref{main1}) and Eq. (\ref{main2})). 

 
 Results from the generalized and traditional Boltzmann collision operators agree to within numerical tolerances. Of course, this is expected since the traditional Boltzmann collision operator is a limiting case of the generalized collision operator. Nevertheless, this comparison also helps to verify the generalized collision operator and our numerical implementation. 

The linear response predicts a slightly larger stopping power than the predictions by the binary collision models. Sources of discrepancy between these approaches include the absence of the velocity-dependent screening (dynamic screening) in the Debye-H\"{u}ckel potential used for modeling the binary collisions, as well as uncertainty in the short-range cut off length (Landau length) used in the linear response theory to avoid the logarithmic divergence caused by neglecting strong nonlinear scattering associated with close collisions~\cite{nicholson1983introduction}. The strengths and weaknesses of these models were previously studied using molecular dynamics simulations \cite{bernstein2019effects,grabowski2013molecular}, and the results shown in Fig.~\ref{fig:plt_gamma} are consistent with these previous studies.
 
\subsection{Strongly magnetized plasma}

\begin{figure} [!htb] 
\centerline{\includegraphics[width = 1.0\columnwidth]{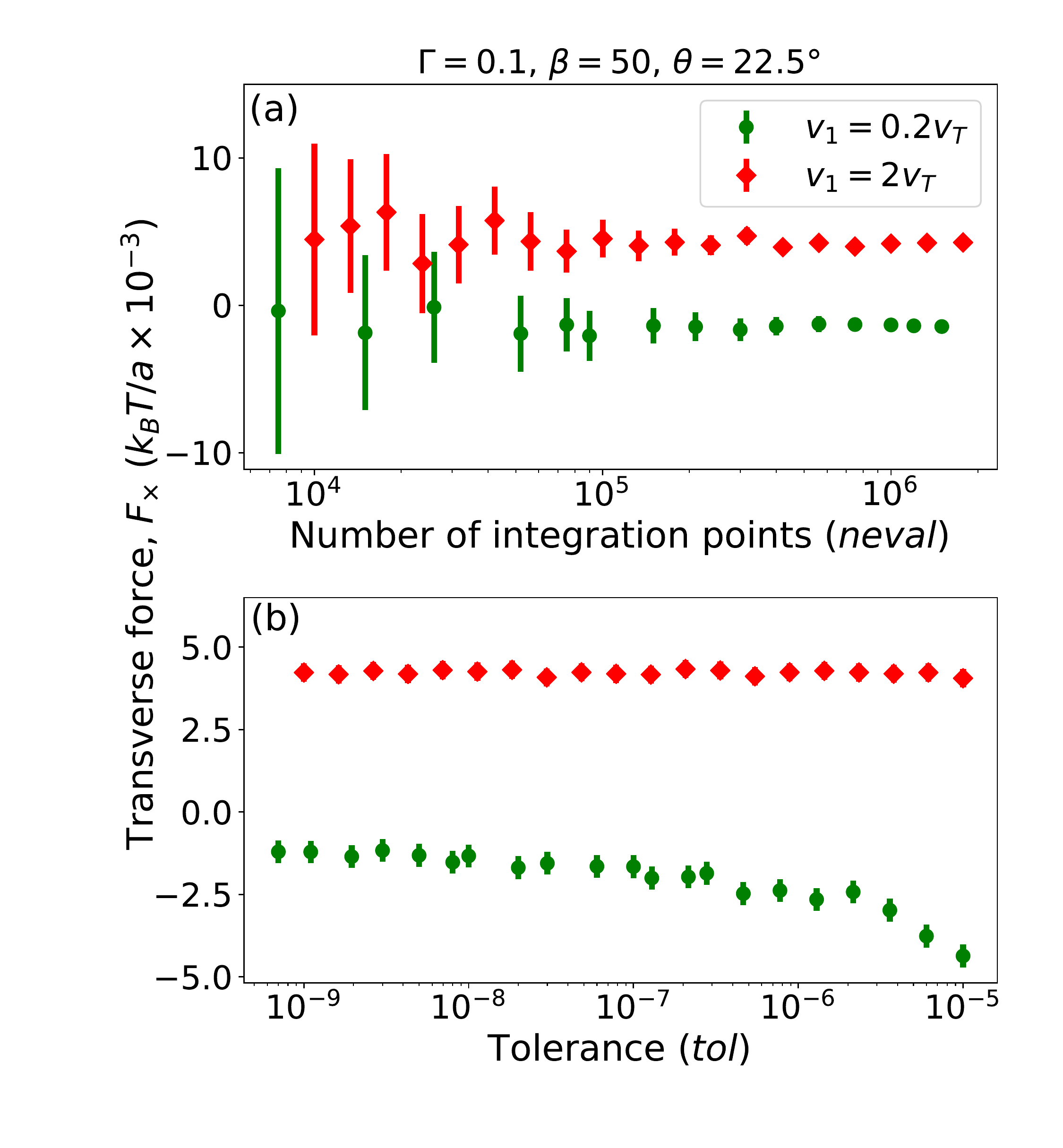}}
\caption {Transverse force ($F_{\times}$) on a projectile in a light background plasma of $\Gamma = 0.1$ and $\beta = 50$ as a function of (a) number of integration points per iteration ($neval$) and (b) tolerance in trajectory calculation ($tol$). The velocity of the projectile makes an angle of $22.5\degree$ with the direction of the magnetic field and initial speed of $v_1=0.2 v_T$ (green circle) and $v_1 = 2 v_T$ (red diamond).}
  \label{fig:conv}
\end{figure}

  In the strongly magnetized regime, the magnetic field influences the collision dynamics, causing the numerical evaluation of the trajectories of colliding particles to become much more computationally expensive. The computational expense was reduced by optimizing the number of integration points per iteration of VEGAS ($neval$) and setting the tolerance for the trajectory calculations ($tol$) in order to achieve a chosen numerical accuracy of the computed friction force coefficients. The tolerance of the trajectory calculation is set by both the relative tolerance and the absolute tolerance, which were taken to be the same. Figure \ref{fig:conv} shows an example convergence test for the transverse force on a projectile in a background plasma of $\Gamma = 0.1 $ and $\beta = 50 $ and having speed $0.2 v_T$ and $2 v_T$. In (a), the tolerance was chosen to be $10^{-8}$ and convergence with respect to the number of integration points was established. In (b), the number of integration points for each iteration was chosen to be $10^6$ and convergence with respect to the tolerance was established. As expected, convergence is obtained as the number of integration points increases, as well as when the tolerance of the trajectory calculation decreases. The number required for convergence was observed to depend on the projectile speed, as well as the $\Gamma$ and $\beta$ parameters. Nevertheless, a tolerance of $10^{-8}$ and number of integration points for each iteration of $10^6$ was sufficient to obtain convergence to less than 1\% throughout the strongly magnetized and extremely magnetized regimes. 

  \begin{figure*} [!htb] 
\centerline{\includegraphics[width = 7.5in]{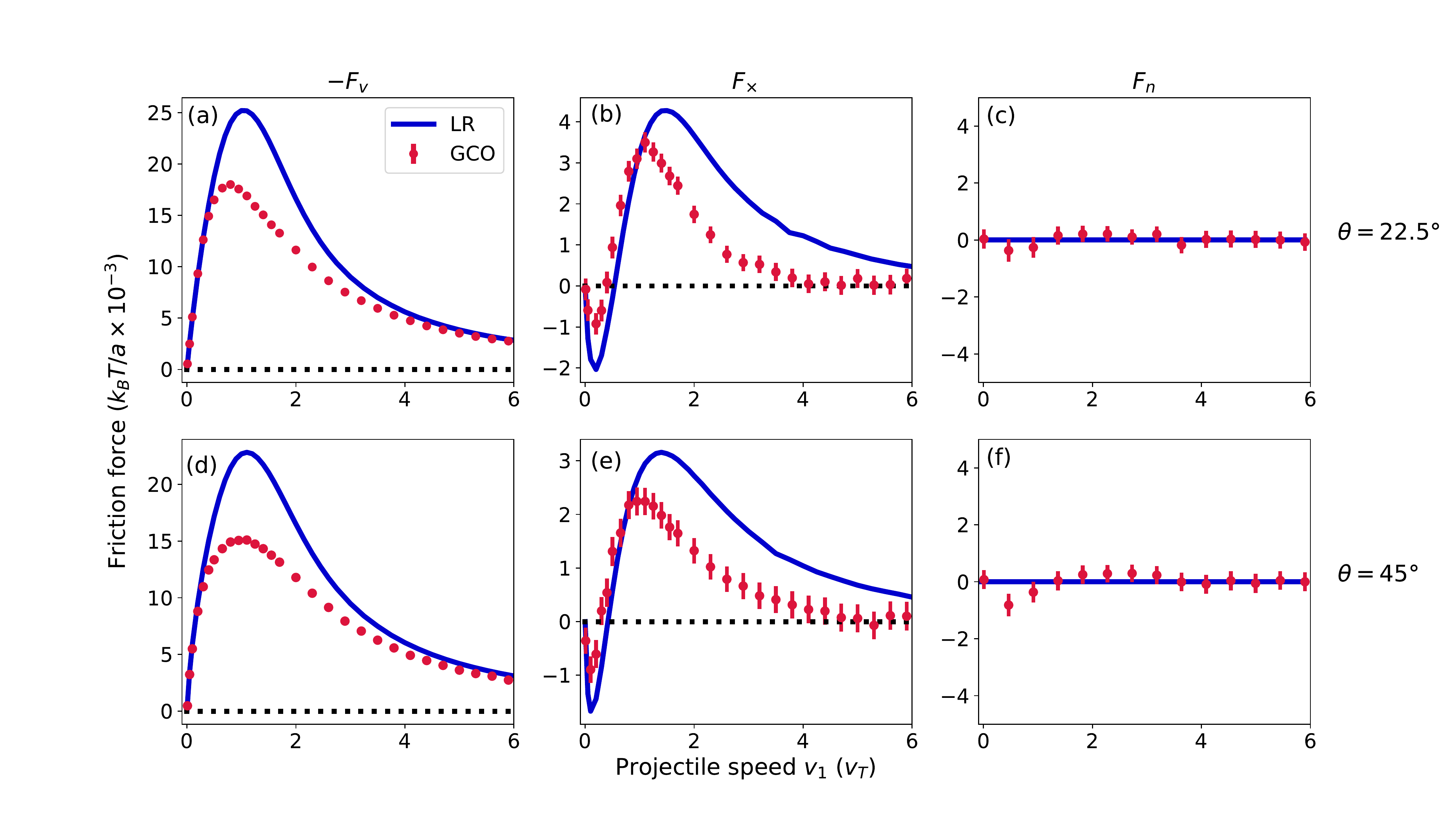}}
\caption {Friction force on a massive projectile ($M=1000$) slowing down on a light background plasma with coupling strength  $\Gamma = 0.1$ and magnetic field strength $\beta =10$ for different initial projectile velocity angles with respect to the magnetic field $\theta =22.5 \degree$ [a, b, and c] and  $\theta = 45 \degree$ [d, e, and f]. The generalized collision operator results (GCO) and linear response theory curve (LR).}
  \label{fig:beta_10}
\end{figure*}
  
Figure \ref{fig:beta_10} shows the friction force curves obtained using the generalized collision operator and the linear response theory for the coupling strength $\Gamma=0.1$ and  the magnetic field strength $\beta=10$ for different orientations of the initial projectile velocity with respect to the magnetic field. Only a qualitative agreement can be reached between the curves from these two theories, because of the shortcomings of the two models that were discussed in the previous subsection.

The main result of this work is shown in panels (b) and (e) of Fig. \ref{fig:beta_10}. This shows that a significant transverse component of the friction force is predicted by the GCO computations. The existence of this component was recently predicted by the linear response approach. Our results demonstrate that this effect is captured in the more complete description from the collision operator of a kinetic theory. It also demonstrates that the effect is captured by the binary collision approach. The two approaches predict qualitatively similar behavior, but have quantitative differences at a similar level to what was observed in the unmagnetized case. This is expected at $\Gamma = 0.1$ due to the uncertainties associated with the screening model, or the short-range cutoff. However, the two approaches would be expected to merge as the coupling strength decreases. 


In linear response theory, the friction force on the projectile is due to the induced electric field associated with the wake generated by the projectile in the background plasma. But in the binary collision theory, the friction force is the net force acting on the projectile from subsequent binary interactions with the background plasma. The linear response theory attributes the origin of the transverse force to the way in which the Lorentz force on the background plasma influences the instantaneously generated wake. In contrast, the generalized collision operator captures the transverse force by accounting for the gyromotion of the background particles while interacting with the projectile. Even though these two are completely different approaches, they both are equally capable of capturing the physics of transverse force in this regime.

On comparing the stopping power curves ($-F_v$) from Fig.~\ref{fig:beta_10} with those for the weakly magnetized regime from Fig.~\ref{fig:plt_gamma}, qualitative changes are observed. The position of the peak shifts to a lower speed, and the magnitude of the force decreases at the Bragg peak, but decays less rapidly with speed. The stopping power is also observed to depend on the orientation of the projectile velocity with respect to the magnetic field. The friction force along the direction of the Lorentz force ($F_n$) is much smaller than either the stopping power or transverse force. Points computed at most velocities are consistent with zero to within the estimated accuracy of the data, but there are a few points at which the computed force appears to be non-zero. This is a qualitative distinction with the predictions of linear response theory and will be studied in greater detail in future work.


\subsection{Extremely magnetized plasma}

\begin{figure*} [!htb] 
\centerline{\includegraphics[width = 7.5in]{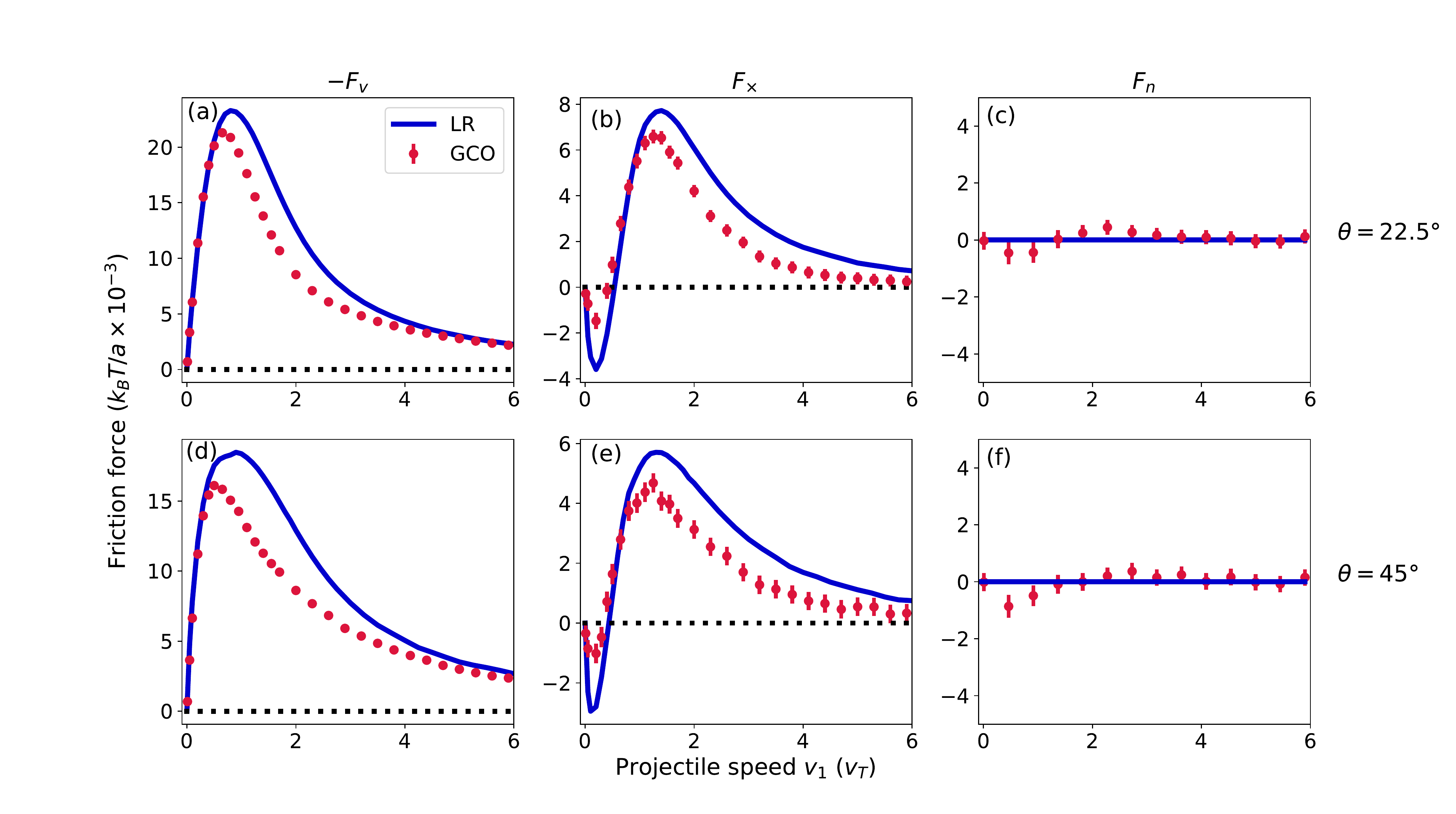}}
\caption {Friction force on a massive projectile ($M=1000$) slowing down on a light background plasma with coupling strength  $\Gamma = 0.1$ and magnetic field strength $\beta =50$ for different initial projectile velocity angles with respect to the magnetic field $\theta =22.5 \degree$ [a, b, and c] and  $\theta = 45 \degree$ [d, e, and f]. The generalized collision operator results (GCO) and linear response theory curve (LR).}
  \label{fig:beta_50}
\end{figure*}
Figure \ref{fig:beta_50} shows the friction force curves obtained using the generalized collision operator and the linear response theory for the coupling strength $\Gamma=0.1$ and  the magnetic field strength $\beta=50$ for different orientations of the initial projectile velocity with respect to the magnetic field. On comparing with the friction force curves in the strongly magnetized regime, the magnitude of the friction force is predicted to increase and the peak of the stopping power curve shift slightly to lower projectile speeds. Similar to the case of strongly magnetized regime, the friction force in the direction of the Lorentz force, $F_n$ is much smaller than either the stopping power or transverse force. 

Linear response theory assumes that the interactions are weak and are small angle collisions. In order to avoid the divergence in the theory caused by the strong large-angle collisions an ad hoc short-range cut off is introduced at the Landau length. These assumptions break down in the extremely magnetized regime. In this regime, the gyroradius is the smallest length scale and the particles are bound to the magnetic field lines. The collisions between the particles are strong and are large-angle collisions. However, the physics of strong interactions are captured by the binary collision theory. This makes the generalized collision operator a strong candidate to understand the physics of the extremely magnetized plasmas. Although the $k_{\max}$ cutoff used in the linear response theory is not expected to apply in the extremely magnetized regime, Fig.~\ref{fig:beta_50} shows a similar level of agreement between linear response theory and the GCO as is observed in the strongly magnetized regime shown in Fig.~\ref{fig:beta_10}. It is unknown if this a fortuitous agreement particular to this combination of $\Gamma$ and $\beta$, or if it will also extend to yet stronger magnetization. 

\section{Discussion}
\begin{figure}[h]  
\centerline{\includegraphics[width = 3.4in]{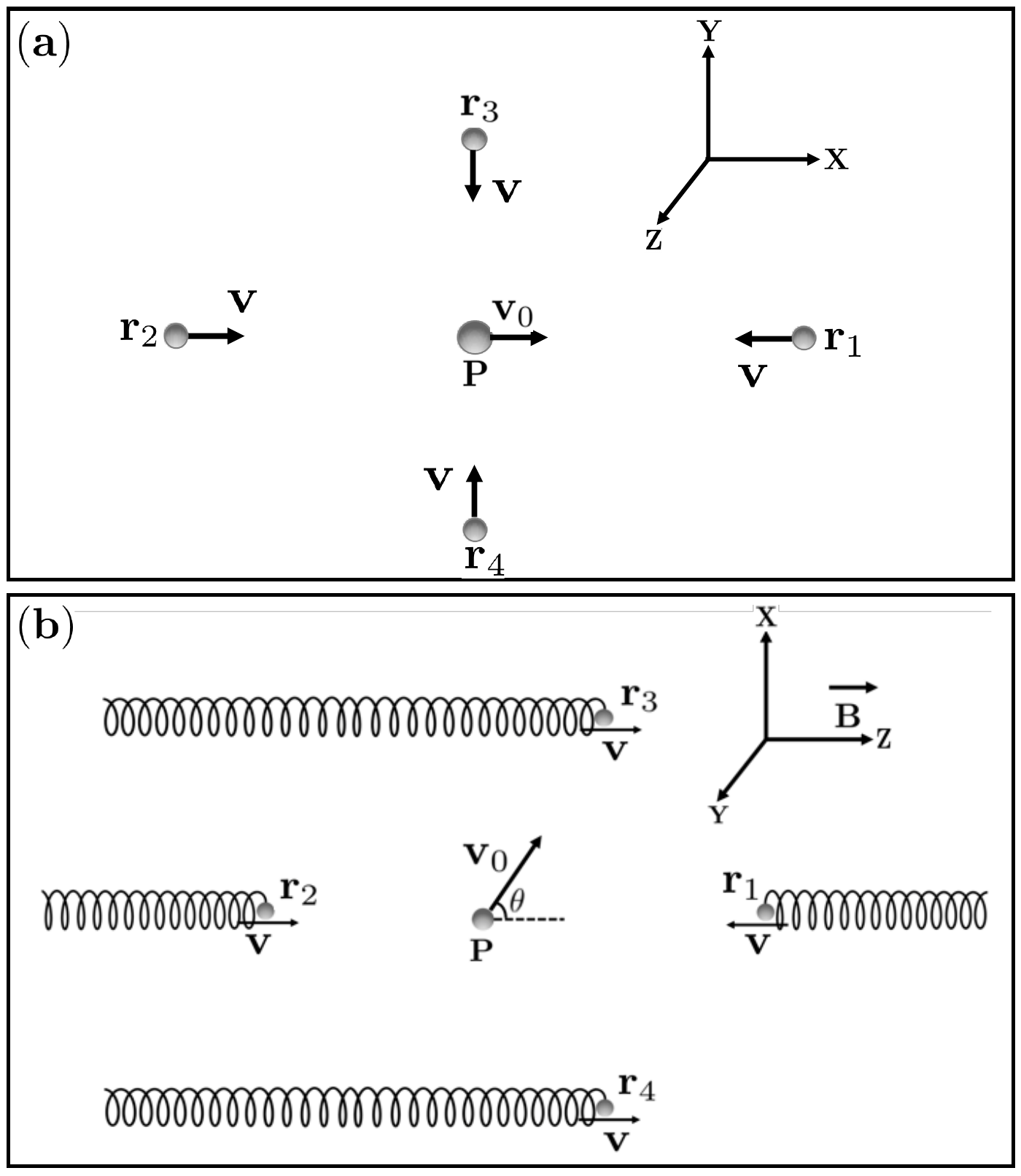}}
\caption{Illustration of collisions of the massive projectile with the background particles - (a) unmagnetized and weakly magnetized (b) strongly magnetized and extremely magnetized transport regimes.}
  \label{fig:discu}
\end{figure}

This section provides a qualitative description of the physical origin of the transverse friction force due to strong magnetization from the binary collision perspective. Binary collision theory calculates the friction force via the change in momentum of the projectile after a sequence of elementary binary collisions with the background plasma particles that includes all possible scattering events. When the gyroradius of the colliding particles are larger than the characteristic scattering length ($\lambda_D$), the influence of the magnetic field during the collisions are negligible. In this case, background plasma particles collide with the projectile from all the directions with equal probability. The net change in momentum from the collisions in all the directions except parallel or antiparallel to the projectile velocity will be zero. For instance, take the projectile velocity to be in the $ +\hat{x}$ direction as shown in panel (a) of Fig. \ref{fig:discu}. Here, the projectile is labeled (P) and the background particles are numbered $r_1$ to $r_4$. The change in momentum of the projectile from a collision with a background particle having velocity $ v\hat{y}$ ($\vc{r}_4$) is canceled by the change in momentum from a collision with a background particle having velocity $ -v \hat{y}$ ($\vc{r}_3$). This leads to the conclusion that there is no transverse component of the friction. In contrast, the change in momentum of the projectile from collisions with the background particle having velocity $- v \hat{x}$ ($\vc{r}_1$) is greater than from a collision with velocity $+v\hat{x}$ ($\vc{r}_2$) because of the larger magnitude of the relative velocity. Thus the friction force is antiparallel to the projectile velocity.

When the plasma is strongly magnetized or extremely magnetized, background particles are bound to the magnetic field lines, effectively making their motion 1-D. This restricts the approach of background particles to the projectile along the $\pm\hat{b}$ direction, breaking the symmetry of particles approaching uniformly from all directions, as in the unmagnetized and weakly magnetized cases. Panel (b) of Fig. \ref{fig:discu} shows the collision of a projectile with four different background particles. The projectile (P) velocity vector makes an angle $\theta$ with respect to the magnetic field and the background particles are numbered $\vc{r}_1$ to $\vc{r}_4$.

Consider the case that the velocity vector of the projectile makes an acute angle to the magnetic field ($0\degree \leq \theta \leq 90\degree$). The net change in momentum of the projectile along the magnetic field direction from collisions with the background plasma particles approaching from the $+\hat{b}$ direction will be higher than that of the collision with background plasma particle approaching from the $-\hat{b}$ direction, resulting in a force anti parallel to the direction of magnetic field. For example consider the collision between particle 1 $(\vc{r}_1)$ and particle 2 ($\vc{r}_2$). Both the particles have equal speed, but the relative velocity is higher for the collision with particle 2, resulting in higher exchange of momentum.

The projectile also experiences a net force in the $-\hat{x}$ direction. This can be understood by comparing collisions between the projectile and particle 3 ($\vc{r}_3$) or particle 4 ($\vc{r}_4$). The projectile experiences more change in momentum from the collision with particle 3 than collision with particle 4 because the projectile is moving towards the particle 3. The conclusion of these arguments is that the presence of the magnetic field breaks the symmetry about the velocity vector, causing there to be both a stopping power component antiparrallel to the velocity and a transverse component perpendicular to the velocity in the plane of $\vc{v}_0$ and $\vc{B}$.
 
The force on the projectile in the $\hat{y}$ direction is expected to be zero because the projectile has no component of the velocity in this direction, other than its gyromotion, and the symmetry of momentum exchange with particles entering from either $\pm \hat{y}$ directions is expected to balance. For instance, the force on the projectile in the $\hat{y}$ direction from a background particle moving along the magnetic field line at $|y|\hat{y}$ is canceled by the force from the background particle moving along the field line at $-|y|\hat{y}$. 

The above discussion considered oblique angles between $\vc{v}_0$ and $\vc{B}$, but when the projectile moves either perpendicular or parallel to the magnetic field, symmetry about the projectile velocity vector is expected to return, and the transverse component of the friction force to vanish. Consider a projectile moving perpendicular to the magnetic field in the $\hat{x}$ direction. In this case, the projectile is not expected to experience any net force in the $\hat{z}$ direction as the momentum exchanged in collisions with background particles approaching from the $+\hat{b}$ and $-\hat{b}$ directions are antisymmetric. In this case, the projectile experiences a force in the $- \hat{x}$ direction only. Thus, a projectile moving perpendicular to the magnetic field only has a stopping power component and no transverse force. Similar arguments of symmetry can be made to understand why there is also no transverse component when the projectile velocity aligns along the magnetic field. Although the solutions in the previous section focused only on oblique angles, these symmetry properties were confirmed, and they have also been shown to hold in both the previous linear response calculations~\cite{lafleur2019transverse} and molecular dynamics simulations~\cite{PRL_David}.



\section{Conclusion}
This work has developed a generalized collision operator applicable to plasmas in the presence of a magnetic field of arbitrary strength. It consists of a 3-D velocity space integral and a 2-D spatial integral on the surface of a collision volume, inside of which the particles interact via the potential of mean force. The size of the collision volume is determined by the range of the potential of mean force. The generalized collision operator incorporates the magnetic field into the collisions by including the Lorentz force acting on the colliding particles in the equations of motion describing the binary collision. The traditional Boltzmann collision operator for the unmagnetized and the weakly magnetized plasma and O'Neil's Boltzmann-like collision operator for the extremely magnetized plasma were obtained from the generalized collision operator by simplifying the collision geometry and equations of motion for the interacting particles in the limits of no magnetic field and high magnetic field, respectively.

The generalized collision operator was applied to compute the friction force acting on a massive projectile moving through the magnetized OCP. The numerical implementation of the generalized collision operator was verified by comparing results from the unmagnetized and weakly magnetized plasma cases with the accepted results of the traditional Boltzmann kinetic theory. The friction force curve for the strongly magnetized plasma was compared with previous results from the linear response theory. The work also extended the computation of the friction force to the extremely magnetized transport regime which was not attainable using the linear response theory. This shows the versatility of the generalized collision operator to obtain the transport properties of the plasmas in all the transport regimes.

By extending the kinetic theory to the transport regimes with the gyroradius smaller than the characteristic scattering length ($\lambda_D$), we now have a theory to understand the basic properties of the plasmas in the strongly magnetized and the extremely magnetized transport regimes. Capturing the influence of the magnetic field on binary collisions has described a new physical effect - the transverse friction force. The transverse force mixes the parallel and perpendicular velocity components of the projectile. How the mixing of these components changes other macroscopic transport properties is unknown. The generalized collision operator is a strong candidate to explore this question.

Many experiments in which strongly magnetized plasmas are found, such as ultracold neutral plasmas~\cite{PhysRevLett.100.235002}, non neutral plasmas~\cite{PhysRevLett.68.317}, and magnetized dusty plasma experiments~\cite{Thomas_2012} exhibit strong Coulomb coupling ($\Gamma>1$). The extension of the generalized collision operator from weak coupling to strong coupling may be attained by using the potential of mean force as the effective interaction potential~\cite{PhysRevLett.110.235001,baalrud2014extending}. This will be explored in future work. 

\section{Data Availability Statement}

The data that support the findings of this study are available from the corresponding author upon reasonable request.

\begin{acknowledgments}
The authors thank Dr. J\'{e}r\^{o}me Daligault for helpful conversations during the development of this work, Dr. Trevor Lafleur for providing the codes used to compute the linear response theory curves and Prof. Peter Lepage for suggestions with implementation of VEGAS.

This material is based upon work supported by the Air Force Office of Scientific Research under Award No. FA9550-16-1-0221, by the U.S. Department of Energy, Office of Fusion Energy Sciences, under Award No. DE-SC0016159, and the National Science Foundation under Grant No. PHY-1453736. 
\end{acknowledgments}

\bibliography{references_1}	

\end{document}